\documentclass[prl,twocolumn,showpacs,preprintnumbers,amsmath,amssymb,floatfix]{revtex4}

\usepackage{graphicx}
\usepackage{dcolumn}
\usepackage{bm}
\usepackage{color}

\begin{document}

\title{Observation of Density Segregation inside Migrating Dunes}
\author{Christopher Groh$^{1}$}
\author{Ingo Rehberg$^{1}$}
\author{Christof A.~Kruelle$^{1,2}$}
\email{christof.kruelle@hs-karlsruhe.de}
\affiliation{$^{1}$Experimentalphysik V, Universit\"{a}t Bayreuth, 
D--95440 Bayreuth, Germany}
\affiliation{$^{2}$Maschinenbau und Mechatronik, 
Hochschule Karlsruhe - Technik und Wirtschaft, D-76133 Karlsruhe, Germany}

\date{\today}

\begin{abstract}
Spatiotemporal patterns in nature, such as ripples or dunes, formed by a 
fluid streaming over a sandy surface show complex behavior despite their 
simple forms. Below the surface, the granular structure of the sand particles 
is subject to self-organization processes, exhibiting such phenomena as 
reverse grading when larger particles are found on top of smaller ones. 
Here we report results of an experimental investigation with downscaled 
model dunes revealing that, if the particles differ not in size but in 
density, the heavier particles, surprisingly, accumulate 
in the central core close to the top
of the dune. This finding contributes to the understanding of sedimentary 
structures found in nature and might be helpful to improve existing 
dating methods for desert dunes.
\end{abstract}


\maketitle

If a fluid streams over an extended area of sand, the grains will 
self-organize by forming complex granular structures like ripples or dunes 
\cite{Bagnold:1941}. The dynamics of these systems is determined by the 
individual fate of the particles \cite{Groh:2010}. In general, agitated 
granular matter is known to show de-mixing whenever particles differ in 
size or density \cite{Kudrolli:2004}, and indeed size segregation is a 
well known feature of ripples and dunes 
\cite{Hunter:1977,Sarre:1990,Tsoar:1990,Fryberger:1992,Anderson:1993,
Julien:1993,Makse:2000,Caps:2002,Wilson:2003,Nickling:2004,Rousseaux:2004,
Jerolmack:2006,Manukyan:2009},
as reviewed by Kleinhans \cite{Kleinhans:2004}.

Already in 1993 Anderson \& Bunas \cite{Anderson:1993} demonstated 
the effect of {\em size} segregation in a migrating dune by modelling
the trajectories of large and small particles with a stochastic cellular automaton.
This kind of particle sorting observed in natural dunes is known to geomorphologists
as \lq reverse grading\rq, where larger particles are more likely to be found
in the upper parts of a dune. 
These coarsened crests occur only for 
fully developed dunes, when their lee slopes are shadowed from impacting
particles.
Then, the trajectory of an individual grain behind the dune's crest
depends strongly on the wind velocity field and its mass.
Larger, and therefore heavier, grains travel in small jumps 
of the order of a few grain diameters, while smaller particles 
are able to leap over the crest far down into the shadow zone.
Consequently, smaller particles end up being buried by larger ones at the toe of a dune.

In addition, Makse \cite{Makse:2000} showed that size segregation 
due to different hopping lengths in the wake of a dune 
competes with a so-called {\em shape} segregation during 
transport and rolling of particles with different roughness
along the dune's surface.
These two different segregation mechanisms in aeolian sand ripples
lead to a so-called
\lq inverse grading climbing ripple lamination\rq\ and 
\lq cross-stratification patterns\rq, 
respectively.
The origin of the cross-stratification structures reminds
to recent experiments of avalanche segregation of mixtures 
of large faceted grains and small rounded grains 
poured in a vertical Hele-Shaw cell 
\cite{Makse:1997,Makse:1999}.

If, however, the particles inside a dune differ not in size or shape but in 
density the question arises whether a similar de-mixing of heavy and light
grains can be found like in other granular systems \cite{Kudrolli:2004}.
Up to now this process of {\em density} segregation during dune migration has 
never been observed experimentally. By using a bi-dense mixture of equally 
sized particles we demonstrate the counter-intuitive effect that the heavier 
particles accumulate 
in the central core close to the top
of the dune. This result provides a clue 
towards an advanced understanding of the sedimentary structures found in nature 
\cite{Allen:1982,Collinson:1982,Bristow:2000}. 
It might be helpful to improve existing dating methods for desert dunes 
\cite{Schuster:2006,Sun:2006} and can illuminate the origin of placers of 
minerals in various sediments \cite{Saxton:2008}.

\begin{figure}[b]
\centering
\includegraphics[width=8cm]{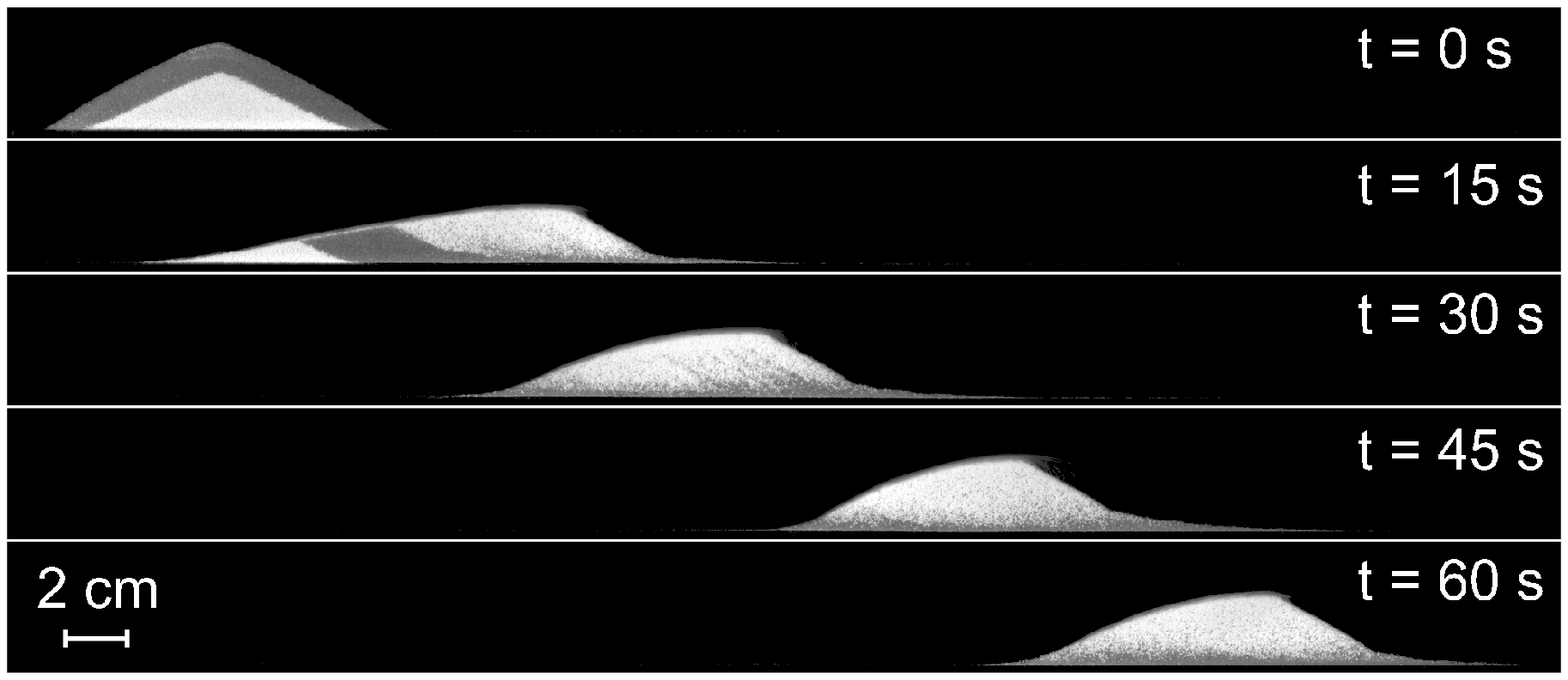}
\caption{Temporal evolution of a bi-dense dune: five time-sequenced snapshots 
showing side views of a developing dune slice composed of glass beads, 
appearing grey, and white ceramic beads. The direction of the 
steady water flow is 
from left to right.}
\label{fig1}
\end{figure}

Our experimental setup mainly consists of a narrow water flow channel, which 
allows the experimental investigation of downsized slices of transverse dunes 
\cite{Groh:2008,Groh:2009}. The width of the channel amounts to 6 mm and the height 
is 6 cm. We measure the fixed unidirectional flow velocity 
of 0.60 m/s with an ultrasonic 
Doppler velocimeter. The corresponding Reynolds number of $Re$ = 36000 is 
calculated with the height of the channel. After the flume is filled with 
distilled water, two different types of beads are poured into the channel. 
One type is made of glass with a density of $\rho_{\rm g}$ = 2.5 $\rm g/cm^{3}$. 
The other ones are white ceramic beads with a density of 
$\rho_{\rm c}$ = 6.0 $\rm g/cm^{3}$. Both particle types have the same mean 
diameter of 0.5 mm in order to exclude size segregation in the form of 
inverse grading \cite{Hunter:1977}. A camera records side views of the 
dunes as shown in Fig.~\ref{fig1}. After suitable image processing one 
can clearly distinguish between the black background, the glass beads 
appearing gray, and white ceramic beads.

\begin{figure}[t]
\centering
\includegraphics[width=7.8cm]{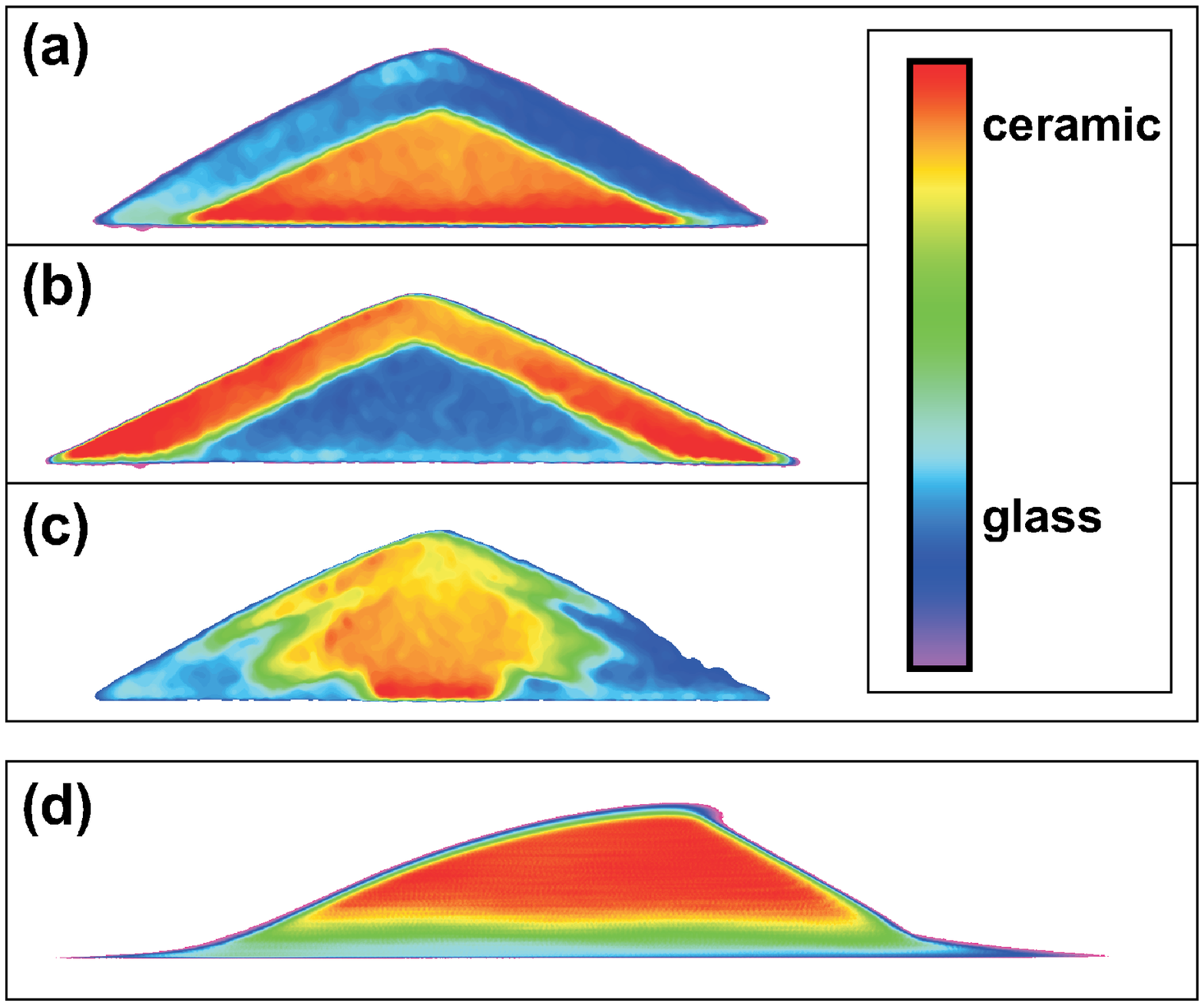}
\caption{(color). Initial preparation and final segregated state for three 
different starting 
configurations: (a) heavy particles on bottom, (b) heavy particles on top, and 
(c) mixed. Panel (d) shows the mass distribution within the common 
steady-state dune slice. The false color code corresponds to the density of the 
particles, i.e.\ red represents heavy and blue light material. A median filter 
smooths the rendering of the beads in (a) to (c). The height of each panel 
corresponds to 4.3 mm in the experiment.}
\label{fig2}
\end{figure}

For the experiment shown in Fig.~\ref{fig1} we choose a fifty-fifty 
by volume mixture 
of glass and ceramic beads. The ceramic beads are poured into the channel 
before the glass beads are filled in. The resulting starting configuration 
is shown in Fig.~\ref{fig1} at $t$ = 0 s. After one minute of continuous 
and unidirectional water flow has elapsed this initial configuration is 
inverted. The lighter glass beads have settled at the bottom of the dune 
while the heavier ceramic ones appear at the top. Moreover, the experiment 
reveals the nature of this paradox effect, which is due to three mechanisms. 
i) The first one is similar to an effect known from agriculture, namely, 
the separation of the wheat from the chaff. In our case the lighter particles 
are dragged by the water flow on longer flights than the heavier ones. 
ii) Further downstream the glass beads are caught in the wake region due 
to the characteristic recirculating flow \cite{Ayrton:1910} and form an 
elongated nose as indicated in Fig.~\ref{fig1}. iii) They provide the carpet 
for the dune to roll over. This explains why the lighter particles always 
end up at the bottom.

To test the universality of the observed segregation effect we prepared two 
other initial starting configurations, namely, the inverse configuration with 
the lighter particles at the bottom (see Fig.~\ref{fig2}(b)), and a mixed 
configuration. The mixed configuration is the attempt to achieve a homogeneous 
mixture of the two types of beads, which succeeds in a beaker, but fails 
during filling this melange into the channel. The different densities lead 
to a configuration as shown in Fig.~\ref{fig2}(c), where the heavier beads 
sediment preferably in the center. The overall mass at each of the three 
experiments remains constant and amounts to $m$ = 22.1 g, as the volume 
fraction of ceramic beads is $\varphi$ = 50 $\%$.

\begin{figure}[t]
\centering
\includegraphics[width=7.8cm]{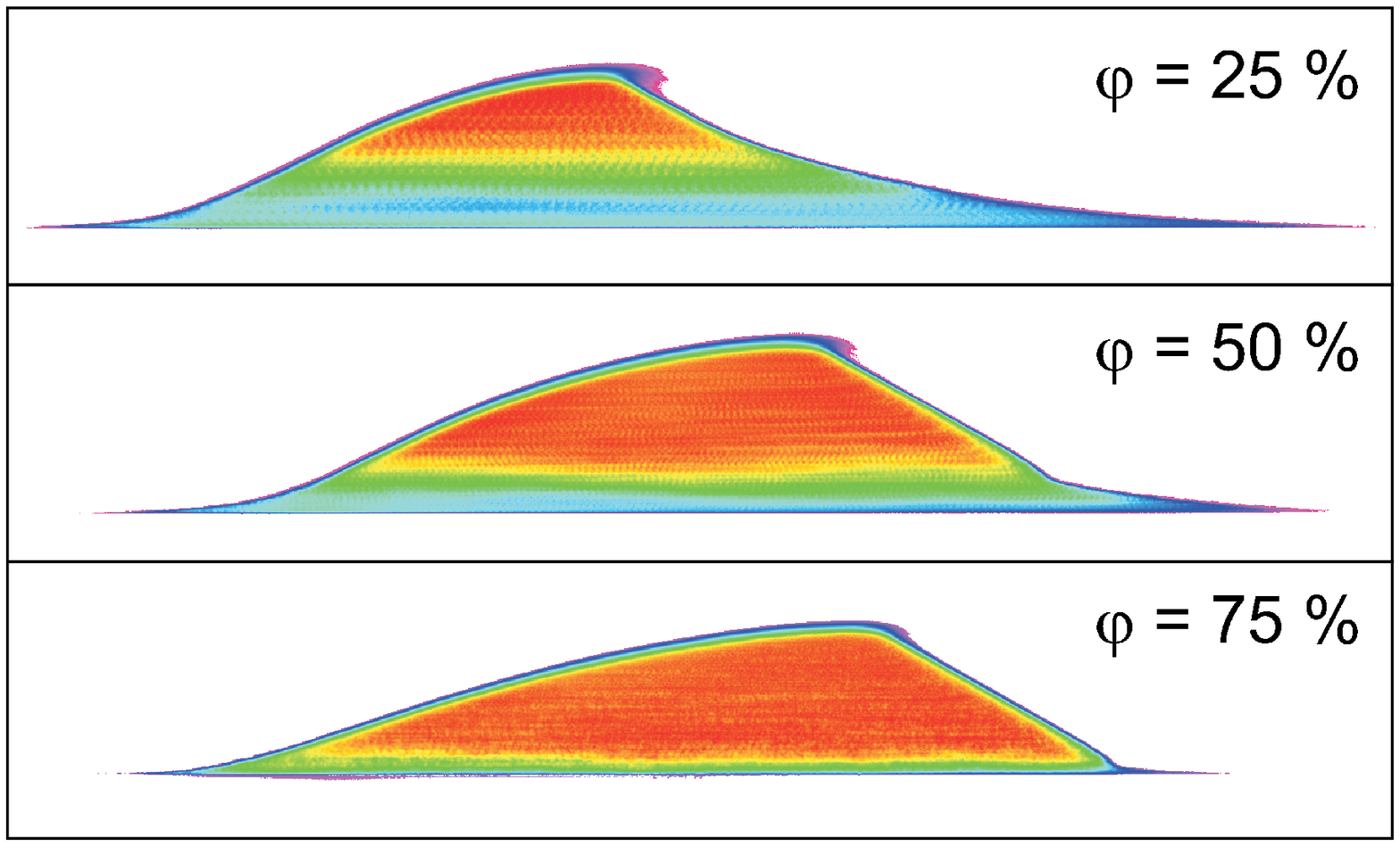}
\caption{(color). Final segregated states for three bi-dense mixtures 
with different volume fractions $\varphi$ of ceramic beads. The color code 
corresponds to the one in Fig.~\ref{fig2}. The height of each panel corresponds 
to 4.3 mm in the experiment.}
\label{fig3}
\end{figure}

Making use of the fact that the three initial configurations achieve one common 
steady state we averaged the images during the last five seconds of each 
measurement. The result is shown in Fig.~\ref{fig2}(d) and represents the 
attractor. The color code indicates the density distribution of the particles, 
which is similar to the one shown in Fig.~\ref{fig1} at $t$ = 60 s.

\begin{figure}[t]
\centering
\includegraphics[width=8cm]{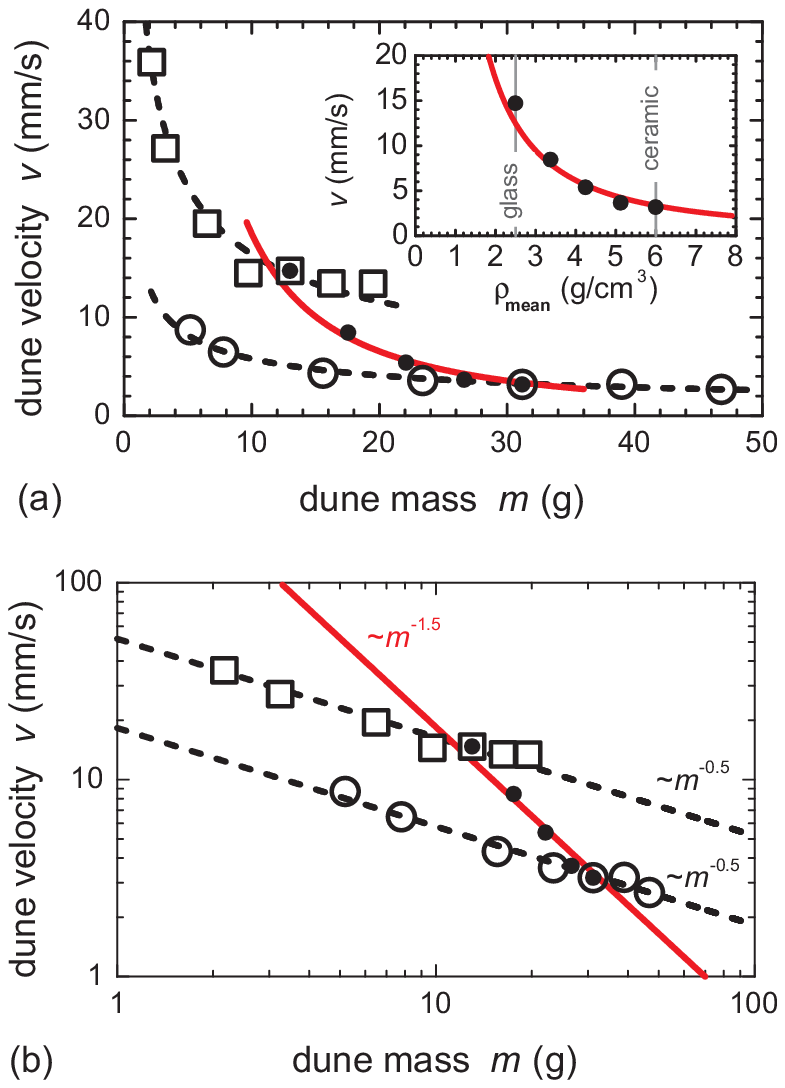}
\caption{(color). 
Linear (a) and double logarithmic plot (b) of the 
migration 
velocity versus dune mass for dunes composed entirely of glass beads 
(squares) or ceramic beads (circles). For bi-dense dunes (full discs) the volume 
is kept constant, but the volume fraction $\varphi$ of ceramic beads is varied, 
which leads to different dune masses. Dashed black lines are power law fits 
with an exponent of -1/2, whereas solid red lines indicate power law fits with 
an exponent of -3/2. In the inset the dune velocity $v$ of bi-dense dunes is 
replotted as a function of the mean particle density $\rho_{\rm mean}$.}
\label{fig4}
\end{figure}

Another option to test the robustness of the observed segregation effect is the 
change of the volume fraction $\varphi$
of dense ceramic beads.
This means that the total size of the 
dune is kept constant, while the percentage of heavy beads is varied. 
Figure \ref{fig3} shows that the heavy beads always end up on top of the dune. 
However, the contours of the dunes differ slightly. At $\varphi$ = 25 $\%$ 
the downstream nose is more pronounced and the slipface formed by the 
ceramic beads is smaller.

To characterize the dynamical behavior Fig.~\ref{fig4} shows the dune 
velocity $v$ as a function of its mass. It is known that 
in the scaling limit
the velocity will 
scale with the square root of the mass, if the dunes consist of only one particle 
species \cite{Andreotti:2002}. 
In particular, it was demonstrated that this scaling holds for at least one 
and a half decades and for three different Reynolds numbers \cite{Groh:2008,Groh:2009}.
The data shown in Fig.~\ref{fig4} for 
pure glass and pure ceramic dunes, respectively, are in agreement with that fact as 
indicated by the dashed black lines, which are fits to 
\mbox{$v\sim m^{-0.5}$}. 

However, for dunes consisting of more particle species of different density we 
expect a scaling according to \mbox{$v\sim \rho_{\rm mean}^{-1.5}$}, 
with the mean density of the mixture given as 
$\rho_{\rm mean}=\rho_{\rm g}+\varphi\cdot(\rho_{\rm c}-\rho_{\rm g})$,
see inset of 
Fig.~\ref{fig4}. This exponent can be explained with the finding of 
Bagnold \cite{Bagnold:1941} that the migration velocity of a steady-state 
dune is inversely proportional to the density $\rho_{\rm mean}$ of its sand. 
Combining both ideas we get $v\sim \rho_{\rm mean}^{-1}\cdot m^{-0.5}$, which 
finally leads to 
\begin{equation}
v\sim m^{-1.5} \qquad {\rm or} \qquad v\sim \rho_{\rm mean}^{-1.5}
\end{equation}
respectively, under the restriction of a constant volume. The 
corresponding fits are shown as red lines in Fig.~\ref{fig4}.

In addition, Fig.\ 5 shows the results of an analysis of the same data as 
displayed in Fig.\ 4. 
Here, the dune velocities for a fixed mass $m$ = 22.1 g (see Fig.\ 1) have been derived
from the prefactors of the scaling laws for the migration velocity $v$ vs.\ dune mass $m$. 
It can be clearly seen that, for a constant mass, the migration velocity scales,
as suggested by Bagnold
\cite{Bagnold:1941} in Chapter 13,
inversely proportional to the average density of the dune's 
particles and thus linearly with the volume of the dune.

\begin{figure}[t]
\centering
\includegraphics[width=8cm]{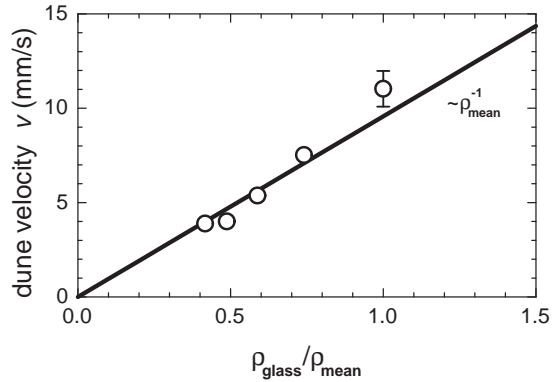}
\caption{
Migration velocity $v$ versus normalized {\em inverse} average density 
$\rho_{\rm glass}/\rho_{\rm mean}$ for dunes composed of glass beads and ceramic beads.
Here the mass $m$ = 22.1 g is kept constant, but the volume fraction $\varphi$ 
of ceramic beads is varied, which leads to different dune volumes. 
The solid line is a linear fit to the data.}
\label{fig5}
\end{figure}

In conclusion, our new measurement technique allows 
to study segregation phenomena both outside and
inside migrating dunes with
high spatial and temporal resolution. This method can even be used for tracking all 
individual particles as demonstrated in 
Ref.\ \cite{Groh:2010}, which revealed the inner 
dynamics of a barchan dune slice.
Our experiment presents the counter-intuitive effect that in bi-dense packings
the heavier grains accumulate at the top of migrating dunes while lighter particles are 
buried at the bottom. 
If the particles' densities were distributed continuously results comparable to the 
inverse grading of polydisperse packings would be expected.

As a side effect we show that the migration velocity of 
bi-dense dunes scales with the mean density of the grains as a power law function 
with an exponent of -3/2. 
This experimental observation raises the question whether the migration velocity 
only depends on $\rho_{\rm mean}$ or whether the {\em distribution} of densities 
also plays a role.

Another issue is the case of transverse dunes. For tackling this task 
new experiments with completely covered floor are needed. There, the dynamics
of a rippled surface developing from an initially flattened bottom has been
studied to a great extent using a single particle species.
For bi-dense packings it is expected that similar segregation
phenomena are observed as for single barchan dunes, with the additional feature,
that lighter particles leaving the lee side of dune will eventually also 
cover the windward side of its downstream neighbor. 

For further examining the underlying mechanisms of the segregation phenomena
observed so far one has to distinguish between size, shape and density segregation
by examining differences in suspended flight behavior between larger/smaller particles 
and denser/less dense particles. 
Based on our experimental findings, it might be speculated that the particle's mass is the 
most relevant parameter.
Here the properties of the driving fluid also play a role: 
e.g.\ for lower Reynolds numbers different mechanisms of entrainment of
particles into the fluid and their successive redeposition on
the dune's surface might result in a different outcome of the segregation patterns.

Loosely speaking, these new insights into the sedimentology of dunes composed of different types of 
sand has the implication, that in a ripple or dune mixed 
of gold and sand, the gold nuggets are likely to be found at the top, 
close to the surface at the crest.

Financial support by the German Science Foundation within Forschergruppe 608
``Nichtlineare Dynamik komplexer Kontinua" through grant Kr1877/3-1 is gratefully
acknowledged.

\end{document}